\documentclass[aps,prapplied,amsmath,amssymb, reprint, superscriptaddress,showpacs]{revtex4-2}
\usepackage{color, graphicx}  
\usepackage{dcolumn}     
\usepackage{bm}                  
\usepackage{amssymb}
\usepackage{lipsum}
\usepackage{latexsym}
\usepackage{amsfonts}
\usepackage{amsmath}
\usepackage{multirow}
\usepackage{ifthen}
\usepackage{hyperref}
\usepackage[dvipsnames]{xcolor}

\usepackage{times}
\begin{document}
	\title{Observation of a $p$-orbital higher-order topological insulator phase in puckered lattice acoustic metamaterials}
	\author{Bing-Quan Wu}
	\affiliation{School of Physical Science and Technology \& Collaborative Innovation Center of Suzhou Nano Science and Technology, Soochow University, 1 Shizi Street, Suzhou, 215006, China}
	\author{Zhi-Kang Lin}
	\email{linzhikangfeynman@163.com}
	\affiliation{School of Physical Science and Technology \& Collaborative Innovation Center of Suzhou Nano Science and Technology, Soochow University, 1 Shizi Street, Suzhou, 215006, China}
	\author{Li-Wei Wang}
 \affiliation{School of Physical Science and Technology \& Collaborative Innovation Center of Suzhou Nano Science and Technology, Soochow University, 1 Shizi Street, Suzhou, 215006, China}
	\author{Jian-Hua Jiang}
 	\email{jhjiang3@ustc.edu.cn}
 \affiliation{School of Biomedical Engineering, Suzhou Institute for advanced research, University of Science and Technology of China, Suzhou, 215123, China}
	\affiliation{School of Physics, University of Science and Technology of China, Hefei, 230026, China}
	\affiliation{School of Physical Science and Technology \& Collaborative Innovation Center of Suzhou Nano Science and Technology, Soochow University, 1 Shizi Street, Suzhou, 215006, China}
 
 \date{\today}

\begin{abstract} 
The puckered lattice geometry, along with $p$-orbitals is often overlooked in the study of topological physics. Here, we investigate the higher-order topology of the $p_{x,y}$-orbital bands in acoustic metamaterials using a simplified two-dimensional phosphorene lattice which possesses a puckered structure. Notably, unlike the $s$-orbital bands in planar lattices, the unique higher-order topology observed here is specific to $p$-orbitals and the puckered geometry due to the unusual hopping patterns induced by them. {Using acoustic pump-probe measurements in metamaterials}, we confirm the emergence of the edge and corner states arising due to the unconventional higher-order topology. We reveal the uniqueness of the higher-order topological physics here via complimentary tight-binding calculations, finite-element simulations, and acoustic experiments. We analyze the underlying physics of the special properties of the edge and corner states in the puckered lattice acoustic metamaterials from the picture of Wannier orbitals. Our work sheds light on the intriguing physics of $p$-orbital topological physics in puckered lattices and acoustic metamaterials which lead to unconventional topological boundary states.
\end{abstract}

\maketitle

\section{Introduction}
The past decades have witnessed tremendous progresses in topological physics and topological materials~\cite{hasan2010colloquium, qi2011topological, wieder2022topological}, bringing in a new era of unconventional physical effects, novel material properties, and potential applications. Recently, higher-order topological insulators~\cite{benalcazar2017quantized,benalcazar2017electric,song2017d,langbehn2017reflection,schindler2018higher,schindler2018higher,benalcazar2019quantization,xie2021higher} hosting lower-dimensional topological boundary states have extended the conventional bulk-edge correspondence to bulk-corner or bulk-hinge correspondence, enriching topological phenomena and topological materials as well as their potential applications. For instance, two-dimensional (2D) higher-order topological insulators such as quadrupole insulators~\cite{benalcazar2017quantized,benalcazar2017electric} support one-dimensional (1D) gapped edge states and zero-dimensional (0D) topological corner states. Topological phases have been widely explored in collective wave dynamics in, e.g., photonic~\cite{lu2014topological,ozawa2019topological,kim2020recent,price2022roadmap,lan2022brief,tang2022topological,wang2022short,zhang2023second,mehrabad2023topological}, acoustic~\cite{ma2019topological,xue2022topological,chen2023various,zhu2023topological}, and circuit metamaterials~\cite{lee2018topolectrical,luo2018topological,dong2021topolectric}, to name just a few, as they can be engineered to simulate topological and quantum phenomena found in electronic systems and even beyond such a paradigm thanks to the nonequilibrium nature of these wave dynamic systems~\cite{zhang2023second}. Furthermore, enabled by metamaterials with fabrication advantages and design diversity as well as various measurement protocols and techniques, classical wave dynamic systems can facilitate the observation and control of topological phenomena.

{To date, most studies on topological phenomena in metamaterials are based on $s$-orbital physics in flat lattices. In contrast to the spherical-like symmetry exhibited by $s$-orbital wavefunctions, $p$-orbital ones possess three orthogonal directionalities in three-dimensional space, namely $p_x$, $p_y$, and $p_z$, as depicted in Fig.~\ref{Fig_1}(a). The phases of $p$ orbitals flip the sign in the opposite direction.} $p$-orbital physics provides a promising approach for band structure and topology engineering~\cite{tokura2000orbital,wu2007flat,wu2008p,wu2008orbital,wirth2011evidence,soltan2012quantum,jacqmin2014direct,milicevic2017orbital,slot2019p,milicevic2019type,gardenier2020p,lu2020orbital,hitomi2021multiorbital,schulz2022photonic,chen2022observation,wang2022valley,zhang2023realization,li2023disentangled,liu2023mono,gao2023orbital,gao2023topological},due to their directionalities. For instance, flat bands can be easily achieved in $p$-orbital systems\cite{wu2007flat,jacqmin2014direct,milicevic2019type,gardenier2020p}. Besides, systems with coexisting $s$- and $p$-orbitals can be used to simulate lattices with intrinsic flux per plaquette and hence give access to quadrupole topological insulators without further engineering~\cite{schulz2022photonic}. $p$-orbitals also give rise to topological phases with enriched orbital degrees of freedom~\cite{gao2023orbital,liu2023mono}. In short, $p$-orbitals can considerably enrich the coupling configurations in lattice systems and hence provides opportunities for the study of topological physics and emergent phenomena in metamaterials. Meanwhile, puckered lattices offer rich geometry beyond planar lattices that exist in nature (such as phospherene) and have intrinsic connection with $p$-orbital physics~\cite{hitomi2021multiorbital}. However, the study of $p$-orbital physics in puckered lattices are still missing in metamaterials research.

In this work, we study the higher-order band topology in a two-dimensional simplified phosphorene lattice~\cite{hitomi2021multiorbital}, which is a puckered analog to the honeycomb lattice. Interestingly, the higher-order topology here is only specific to $p$-orbitals due to their unique hopping configurations, whereas the $s$-orbitals give trivial band topology. Besides, in contrast to the $s$-orbital bands in planar honeycomb lattices, we observe unconventional armchair edge states and corner states within a sizable band gap due to the rich $p$-orbital physics in the puckered lattice. Based on the theory of band representations~\cite{bradlyn2017topological,cano2021band}, the $p$-orbital physics in phosphorene lattices falls into the topological class of obstructed atomic insulators with Wannier orbitals located at each bond centers, which explains the underlying higher-order topology revealed in both the tight-binding (TB) calculations and the acoustic simulations.{The theoretical predictions are verified in experiments by using various acoustic pump-probe configurations. The acoustic pump-probe technique involves the utilization of a tiny speaker as the pump source, while simultaneously employing a microphone to probe the acoustic responses at specific locations within acoustic crystals}.

This paper is organized as follows. In Sec.~\ref{aa}, we recall in brief the simplified phosphorene lattice which is a multiorbital TB model, and reveal the existence of the localized edge and corner states induced by $p$-orbitals, The analysis of the topological origin is based on the band representations and Wannier orbital centers. Then in Sec.~\ref{bb} we design in simulation the acoustic analog of the phosphorene lattice and confirm its consistency to the TB model through its band structures for bulk, edge, and corners as well as their acoustic wavefunctions. Furthermore, in Sec.~\ref{cc}, using air-borne acoustic metamaterial as a platform, we experimentally validate the edge and corner states by {acoustic pump-probe} measurements. Finally, a summary is concluded in Sec.~\ref{dd}.

\section{Results}
\subsection{Multi-orbital tight-binding model}\label{aa}
\begin{figure*}[t]
	\centering
\includegraphics[width=2\columnwidth]{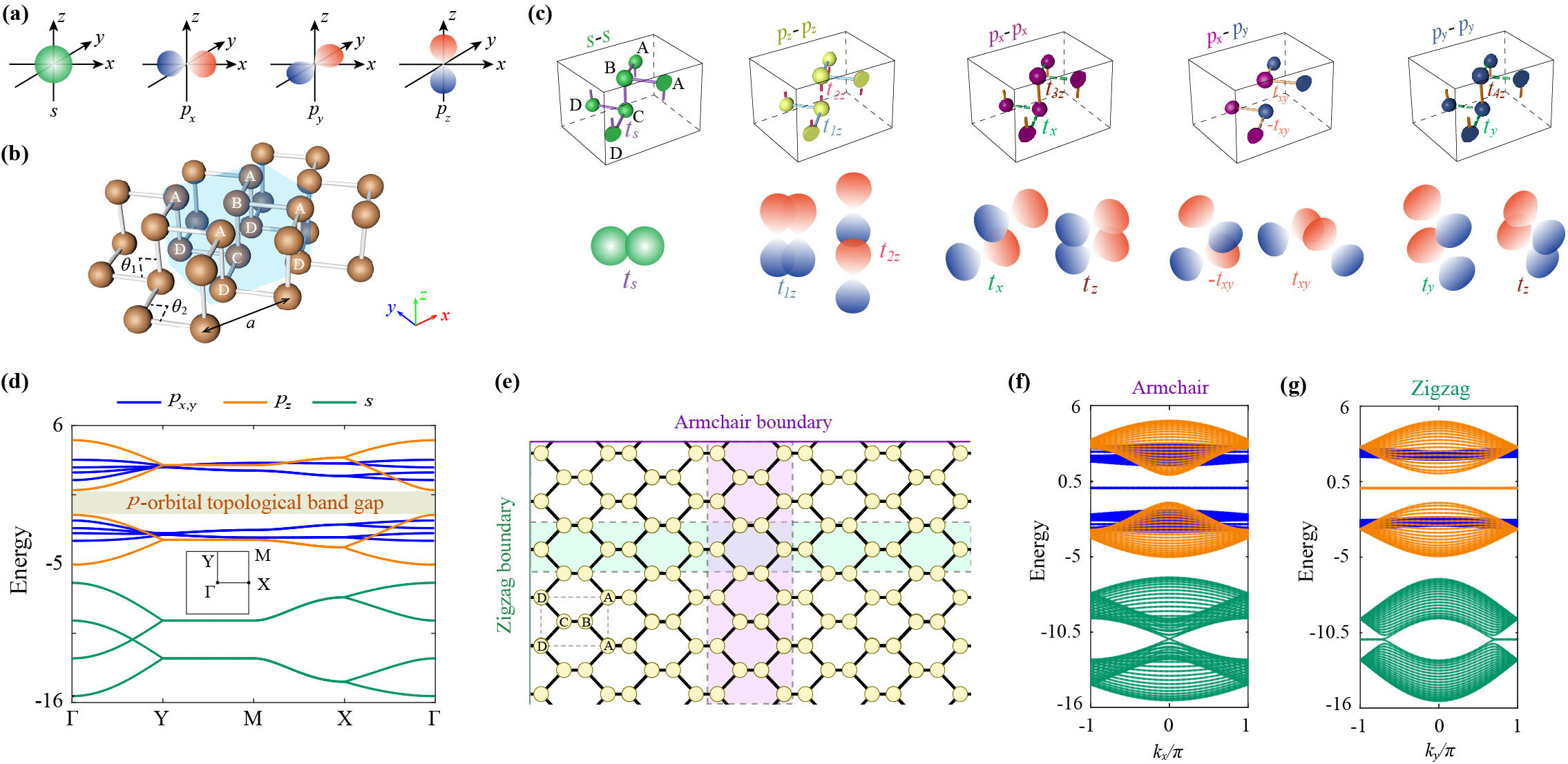}
	\caption[figure_1]{(a) Illustration of $s$-, $p_x$-, $p_y$- and $p_z$-orbitals. (b) Illustration of the tight-binding model for the phosphorene lattice. The unit cell indicated by the transparent blue zone consists of four atomic sites, as denoted by A, B, C, and D, respectively. We choose $\theta_1=\theta_2=90^\circ$ for simplicity. The lattice constants along both the $x$- and $y$-directions are set to unity. (c) The upper panels show five elementary hopping configurations associated with the $s$-, $p_{x,y}$- and $p_{z}$-orbitals. Distinct orbitals and hoppings are represented by different colors. The dashed lines represent negative hoppings which are unique to $p$-orbitals. The lower panels schematically visualize the corresponding hoppings among different orbitals. (d) Band structure of the simplified phosphorene lattice in (b). In the calculation, the hopping parameters labeled in (c) are set as $t_s=-1.5$, $t_{1z}=-1$, $t_{2z}=3$, $t_x=t_y=1.2$, $t_z=-0.3$, and $t_{xy}=-1.3$. The onsite energy of the $s$-, $p_x$-, $p_y$-, and $p_z$-orbitals are chosen as, $\epsilon_s=-11$, $\epsilon_x=0.5$, $\epsilon_y=-0.5$, and $\epsilon_z=0$, respectively. The green, blue, and orange curves represent energy bands originating from the $s$-, $p_{x,y}$- and $p_{z}$-orbitals, respectively. The band gap developed in the $p_{x,y}$-orbital bands is the topological band gap of concern in this work. (e) Illustration of the armchair and zigzag types of edge truncations. The outlined green (purple) region denotes the semi-infinite ribbon cell for calculating the zigzag- (armchair-) type projected band structures. (f) and (g) Armchair- and zigzag-type projected band structures calculated from the ribbon supercells in (e) with a total number of 62 and 60 atomic sites, respectively. }
	\label{Fig_1}                                                       
\end{figure*}

To investigate the intriguing interplay between the puckered lattice geometry, the $p$-orbital physics, and the higher-order band topology, we consider the phosphorene lattice as a prototype puckered lattice in this study. We consider energy bands developed from the $s$- and $p_{x,y,z}$-orbitals. As illustrated in Fig.~\ref{Fig_1}(b), the original phosphorene lattice resembles a nonplanar honeycomb lattice with bond angles $\theta_1\approx103^\circ$ and $\theta_2\approx98^\circ$, which is essentially a monolayer but with a puckered geometry, pertaining to the layer symmetry group $Pman$~\cite{aroyo2006bilbao}. The unit cell is depicted by the transparent blue region, consisting of four inequivalent atomic sites, labeled separately as A, B, C, and D in Fig.~\ref{Fig_1}(b). Only the nearest-neighbor hoppings (indicated by the grey bonds) are considered here, which adequately describe the relevant band topology.

{We then consider a simplified model with $\theta_1=\theta_2=90^\circ$, which preserves the same topological classification as the original phosphorene lattice due to this ``orthogonalization''
process that does not alter any inherent spatial symmetry. Furthermore, the simplified lattice facilitates the simulation calculation and the preparation of experimental samples.   
Subsequently, we investigate the simplified phosphorene lattice and take the $s$- and $p_{x,y,z}$-orbitals into account at each atomic site,} which results in a total of 16 orbitals in one unit cell and a $16\times16$ Hamiltonian in momentum space.  The hoppings between the $s$- and $p$-orbitals can be neglected as their onsite energy difference is considerably large. Furthermore, the hoppings between $p_{x,y}$- and $p_z$-orbitals vanish due to orthogonality constraints. With these considerations, there are in total five distinct elementary hopping configurations for the four types of orbitals, as illustrated in Fig. ~\ref{Fig_1}(c). The hopping parameters are denoted by $t_s$, $t_x$, $t_{xy}$, $t_y$, $t_z$, $t_{1z}$, and $t_{2z}$, respectively. Notably, here due to the $p_{x,y}$-orbital geometry, both positive (solid lines) and negative (dashed lines) hoppings exist in the model.

With these considerations and settings, the Hamiltonian can be written in the following block diagonal form,
 \begin{align}
	\mathcal{H}(\boldsymbol{k})  =diag &\left[\begin{array}{ccc}
		\mathcal{H}_s(\boldsymbol{k}), & \mathcal{H}_{xy}(\boldsymbol{k}), & 
  \mathcal{H}_z(\boldsymbol{k})\\
	\end{array}\right],
\end{align}
where the subscripts $s$, $xy$, $z$ stand for $s$-, $p_{x,y}$-, and $p_z$-orbitals, respectively. The $4\times4$ block Hamiltonian $\mathcal{H}_{i}(\boldsymbol{k}) \enspace (i=s, z)$ can be written in the basis of the A, B, C, and D sites as following
\begin{align}
	\mathcal{H}_i(\boldsymbol{k})  = &\left[\begin{array}{cccc}
		\epsilon_i & h(t_{1i}) & 0 & t_{2i} \\
		h^\ast(t_{1i}) & \epsilon_i  & t_{2i} & 0 \\
		0 & t_{2i} & \epsilon_i & h(t_{1i})  \\
  t_{2i} & 0 & h^\ast(t_{1i}) & \epsilon_i  \\
	\end{array}\right] ,
\end{align} 
where
\begin{align}
h(t_{1i})=t_{1i}e^{i(-\frac{1}{2} {k_x}+\frac{1}{2}{k_y})}+t_{1i}e^{i(-\frac{1}{2}{k_x}-\frac{1}{2}{k_y})} .
\end{align}
Here, the lattice constant $a$ is set to unity in the TB model. The hopping parameters are chosen as follows: $t_{1s}=t_{2s}=t_s=-1.5$, $t_{1z}=-1$, and $t_{2z}=3$. $\epsilon_s=-11$ and $\epsilon_z=0$ are the onsite energy for the $s$- and $p_z$-orbitals, respectively. $k_x$ and $k_y$ are the components of the 2D wavevector $\boldsymbol{k}$ along the $x$- and $y$-directions, respectively.

Meanwhile, the $8\times8$ block Hamiltonian $\mathcal{H}_{xy}(\boldsymbol{k})$ can be explicitly written as
\begin{widetext}
\begin{align*}
	\mathcal{H}_{xy}(\boldsymbol{k})  = &\left[\begin{array}{cccccccc}
		\epsilon_x & h(t_{x}) & 0 & t_{z} & 0 & h(t_{xy}) & 0 & 0 \\
		h^\ast(t_{x}) & \epsilon_x & t_z & 0 & h^\ast(t_{xy}) & 0 & 0 & 0 \\
  0 & t_z & \epsilon_x & h(t_{x}) & 0 & 0 & 0 & h(t_{xy}) \\
  t_z & 0 & h^\ast(t_{x}) & \epsilon_x & 0 & 0 & h^\ast(t_{xy}) & 0 \\
  0 & h(t_{xy}) & 0 & 0 & \epsilon_y & h(t_{y}) & 0 & t_z \\
  h^\ast(t_{xy}) & 0 & 0 & 0 & h^\ast(t_{y}) & \epsilon_y & t_z & 0 \\
  0 & 0 & 0 & h(t_{xy}) & 0 & t_z & \epsilon_y & h(t_{y}) \\
0 & 0 &	 h^\ast(t_{xy}) & 0 & t_z & 0 & h^\ast(t_{y}) & \epsilon_y\\	
	\end{array}\right] 
\end{align*}
\end{widetext}
with
\begin{align}
h(t_{i})=t_{i}e^{i(-\frac{1}{2} {k_x}+\frac{1}{2}{k_y})}+t_{i}e^{i(-\frac{1}{2}{k_x}-\frac{1}{2}{k_y})}, i=x, y
\end{align}
and
\begin{align}
h(t_{xy})=t_{xy}e^{i(-\frac{1}{2} {kx}+\frac{1}{2}{k_y})}-t_{xy}e^{i(-\frac{1}{2}{k_x}-\frac{1}{2}{k_y})} .
\end{align}
Here, $t_x=t_y=1.2$ and $t_z=-0.3$ are the hoppings along the $x$, $y$, and $z$ directions, separately. These are the hoppings within the $p_x$- or $p_y$-orbital and they are identical for the two types of orbitals. $t_{xy}=-1.3$ is the hopping between the $p_x$- and $p_y$-orbitals. The corresponding onsite energy of the $p_x$- and $p_y$-orbitals are $\epsilon_x=0.3$ and $\epsilon_y=-0.3$, respectively. By diagonalizing the $16\times16$ Hamiltonian $\mathcal{H}(\boldsymbol{k})$, we obtain the TB band structure shown in Fig.~\ref{Fig_1}(d). Here, the blue curves represent the energy bands originating from the $p_{x}$- and $p_{y}$-orbitals, while the green and orange curves represent the energy bands derived from the $s$- and $p_z$-orbitals, respectively. We notice that the energy bands developed from the $p_{x}$- and $p_{y}$-orbitals exhibit a sizable band gap around energy zero which is the band gap of concern in this work. We emphasize that the puckered lattice geometry plays a crucial role in forming the sizable $p_{x,y}$-orbital band gap which, instead, is absent in the planar honeycomb lattice geometry, although the two lattices share many similar properties.

\begin{figure*}[t]
	\centering
	\includegraphics[width=2\columnwidth]{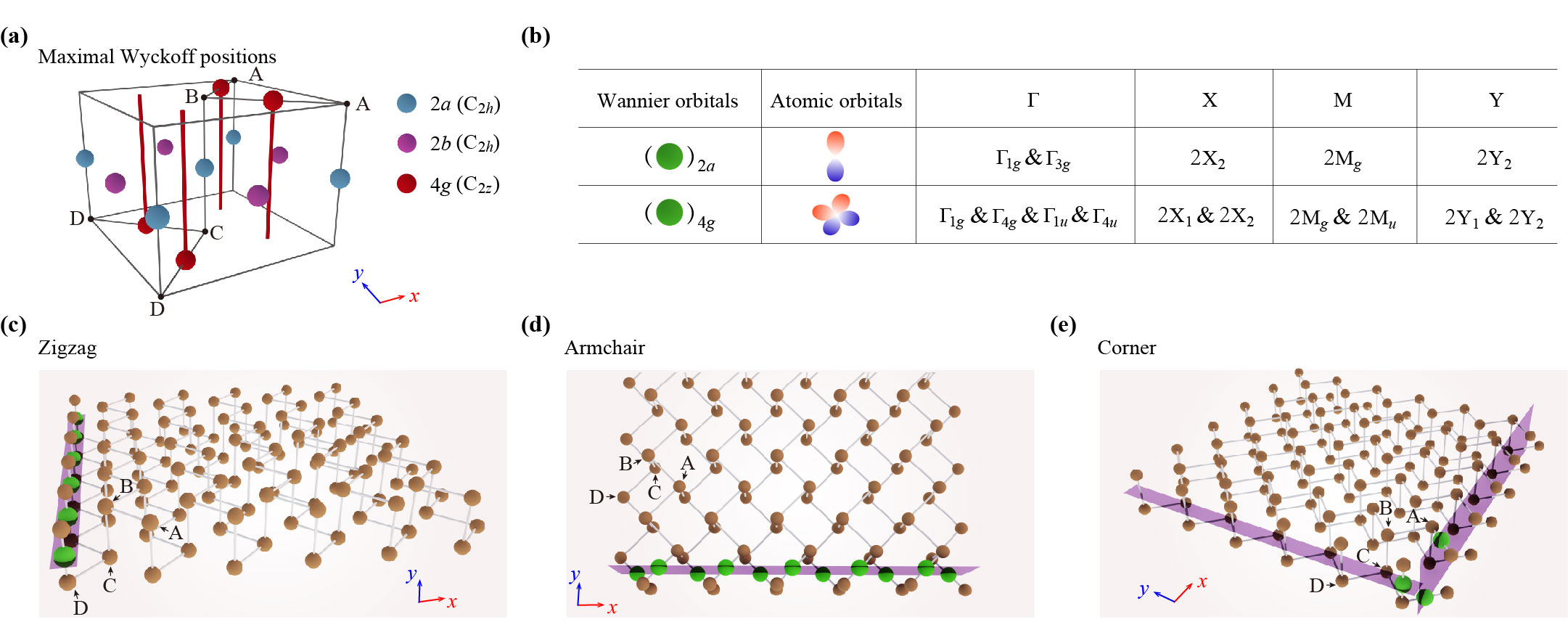}
	\caption{(a) Illustration of maximal Wyckoff positions $2a$, $2b$ and $4g$ for the layer group $Pman$, of which the site-group symmetries are $C_{2h}$, $C_{2h}$ and $C_{2z}$, respectively. Four $4g$ positions are movable along the red lines but subject to the $C_{2y}$ and $M_{y}$. (b) Band representations of the lower $p$-orbital bands. The second and third rows denote the band representations for the $p_z$- and $p_{x,y}$-orbital bands, respectively. The first column represents the corresponding real-space  $s$-Wannier orbitals localized at the maximal Wyckoff positions $2a$ and $4g$. The third to sixth columns include the symmetry representations at high-symmetry points $\Gamma$, $X$, $Y$, and $M$. The labels of these symmetry representations can be referred to Refs.~\cite{hitomi2021multiorbital,aroyo2006bilbao}. (c)-(e) Schematic diagrams of the truncated Wanier orbitals at the armchair edges (c), zigzag edges (d), and corners (e). The $s$-Wannier orbitals, as highlighted by green spheres, are cut through by translucent purple ribbons, which indicates the emergence of edge and corner states.}\label{Fig_2} 
\end{figure*}

We now study the bulk-edge responses in our system. Similar to the honeycomb lattice, the phosphorene lattice hosts two prototypes of edge terminations: the armchair and zigzag edges, as schematically sketched in Fig.~\ref{Fig_1}(e) where we flattened the geometry of the phosphorene lattice to illustrate the edge geometry, particularly to illustrate the difference between the zigzag and armchair edges. The corresponding projected band structures, calculated from the ribbon-shaped supercells illustrated in Fig.~\ref{Fig_1}(e), are shown in Figs.~\ref{Fig_1}(f) and~\ref{Fig_1}(g). We find that for the $s$-orbital bands, the edge states here are similar to the edge states in the planar honeycomb lattice, revealing that the puckered lattice geometry has no effect on the $s$-orbital bands. Intriguingly, for both edge terminations the $p$-orbital band gaps host topological flat-band edge states with flat dispersion. Here, the armchair and zigzag edge states are of the $p_{x,y}$- and $p_z$-orbital nature, separately.

The appearance of flat-band zigzag edge states in the $p_z$-orbital band gap still has its connection with the renowned edge property in the honeycomb lattice~\cite{kohmoto2007zero}. In contrast, the emergence of the flat-band armchair edge states in the $p_{x,y}$-orbital band gap is an unconventional feature due to the $p_{x,y}$-orbital physics. Moreover, calculations for finite-sized lattices indicate the emergence of degenerate edge and corner states sharing the same energy, which is an intriguing manifestation of the higher-order topology. {The degeneration and flat dispersion arise from the intrinsic chiral symmetry in $p_{x,y}$-orbital bands. The chiral symmetry operator here is defined as $\Gamma= $diag$(1, -1, 1, -1,1, -1, 1, -1)$ satisfying 
\begin{align}
\Gamma [\mathcal{H}_{xy}(\boldsymbol{k})-(\epsilon_x+\epsilon_y)/2] \Gamma=-[\mathcal{H}_{xy}(\boldsymbol{k})-(\epsilon_x+\epsilon_y)/2],
\end{align}
which pins all $\boldsymbol{k}$-dependent armchair edge states and the emergent topological corner states at the energy level of $(\epsilon_x+\epsilon_y)/2$.}

{To characterize the higher-order band topology, it is customary to first calculate the bulk polarization and the corner charge. The bulk polarization ${\boldsymbol P}=(P_x, P_y)$ can be derived from the parity-inversion of Bloch states by~\cite{Liufeng}
\begin{align}
P_x=\frac{1}{2}q_x\mod 1, P_y=\frac{1}{2}q_y\mod 1,
\end{align}
where $(-1)^{q_x}=\Pi_j\xi_j(\Gamma)\xi_j(X)$ and $(-1)^{q_y}=\Pi_j\xi_j(\Gamma)\xi_j(Y)$, $\xi(K)$ denotes the  parity-inversion eigenvalue of Bloch wavefunctions at high-symmetry points $K$, that is $+1$ for the even parity and $-1$ for the odd parity. $j$ is the band index below the band gap. Based on the symmetry representations labeled in Fig. ~\ref{Fig_2}(b), where $\Gamma_{1g}$, $\Gamma_{3g}$, $X_1$, and $Y_1$ are of even parities, while $\Gamma_{1u}$, $\Gamma_{4u}$, $X_2$, and $Y_2$ denote odd parities, we obtain the bulk polarization ${\boldsymbol P}=(0,0)$ for both the $p_z$- and $p_{x,y}$-orbital bands. We further calculate the corner charges. Due to the absence of rotational symmetry in the phosphorene lattice, the applicability of topological indices for calculating corner charges as described in Ref.~\cite{benalcazar2019quantization} is limited in this paper. We instead, employ $Q=\Sigma_i \Sigma_j \rho(i,j) ~mod~1$, to directly derive the corner charges~\cite{corneranomaly}, where $\rho(i,j)$ represents the local density of states for a finite-sized system, with $i$ denoting the bulk band index below the band gap and $j$ referring to the four sites in the corner unit cell. Unusually, the corner charges for both the $p_z$- and $p_{x,y}$-orbital bands almost vanish as well. We remark that the nonzero bulk polarizations or fractional corner charges are not the necessary and sufficient conditions for the emergence of topological edge and corner states. A more essential condition is the filling anomaly arising from the off-centered Wannier orbitals~\cite{benalcazar2019quantization}.}

We hereafter study the properties of the $p$-orbital bands using the Wannier orbital configurations in real space and the elementary band representation (EBR) analysis~\cite{bradlyn2017topological,elcoro2017double,cano2021band}. The Wannier orbital picture is more intuitive and straightforward, while the EBRs are powerful tools for the analysis of the band topology. We first figure out the maximal Wyckoff positions in our system which are the high-symmetry positions in a real-space unit cell. The maximal Wyckoff positions for the layer group $Pman$ are $2a$, $2b$, and $4g$, as illustrated in Fig. ~\ref{Fig_2}(a), enjoying the site symmetries $C_{2h}$, $C_{2h}$, and $C_{2z}$, respectively~\cite{aroyo2006bilbao1}. Interestingly, the $2a$, $2b$, and $4g$ Wyckoff positions are at the centers of the hopping bonds AB, BC, and CD, separately. The real-space Wannier orbitals can be derived from the symmetry representations of Bloch wavefunctions at high-symmetry points in the wavevector space via the powerful tool of the EBRs in topological quantum chemistry~\cite{elcoro2017double}. The symmetry representations of the lower six $p$-orbital Bloch bands at the $\Gamma$, $X$, $Y$, and $M$ momentum points are presented in Fig. ~\ref{Fig_2}(b). One can refer to Refs.~\cite{hitomi2021multiorbital,aroyo2006bilbao} for the symmetry representation labels at these high-symmetry points in the Brillouin zone.

By retrospecting the EBRs of the layer group $Pman$~\cite{elcoro2017double}, it is found that the real-space Wannier orbitals of the phosphorene lattice for the $p_z$- and $p_{x,y}$-bands below the band gap of interest are $s$-like Wannier orbitals located at the $2a$ and $4g$ positions, respectively, which indicates that the topological band gap of concern gives an obstructed atomic insulator phase (i.e., materials with the Wannier orbital centers deviating from the atomic positions)\cite{cano2021band,xu2021filling,bradlyn2017topological}. Specifically, the Wannier orbitals of $p_z$-bands reside at the center of the vertical bonds BC and AD, since the $p_z$-orbitals contribute to the formation of $\sigma$ bonds along the $z$-direction. In contrast, the Wannier orbitals for the $p_{x,y}$-bands below the band gap of concern are located at the centers of the in-plane bonds AB and CD.

Topological boundary states can emerge in finite-sized systems when the edge and corner boundaries cut through the centers of the Wannier orbitals. Here, the zigzag edge boundaries cut through the Wannier orbitals centered at the $2a$ Wyckoff position associated with the $p_z$ bands, as shown in Fig.~\ref{Fig_2}(c). Therefore, the $p_z$-band gap support the edge states at the zigzag edge boundaries. In comparison, the armchair edge boundaries cut through the Wannier orbitals centered at the $4g$ Wyckoff position associated with the $p_{x,y}$-bands below the band gap, as depicted in Fig.~\ref{Fig_2}(d). Thus, the armchair edge states appearing in the topological band gap are induced by $p_{x,y}$-orbital bands.

The corner states appearing in the topological band gap can be analyzed similarly. For the corner geometry and the Wannier orbital configuration illustrated in Fig.~\ref{Fig_2}(e), the corner boundary cuts both the $2a$ and $4g$ Wyckoff positions. Therefore, both the $p_z$- and $p_{x,y}$-bands contribute to the emergence of the corner states. We focus only on the topological armchair edge states and corner states induced by $p_{x,y}$-orbital bands in the following, which are unique to the puckered phosphorene lattice but instead absent in the planar honeycomb-like lattices.

\begin{figure*}[t]
	\centering
	\includegraphics[width=2\columnwidth]{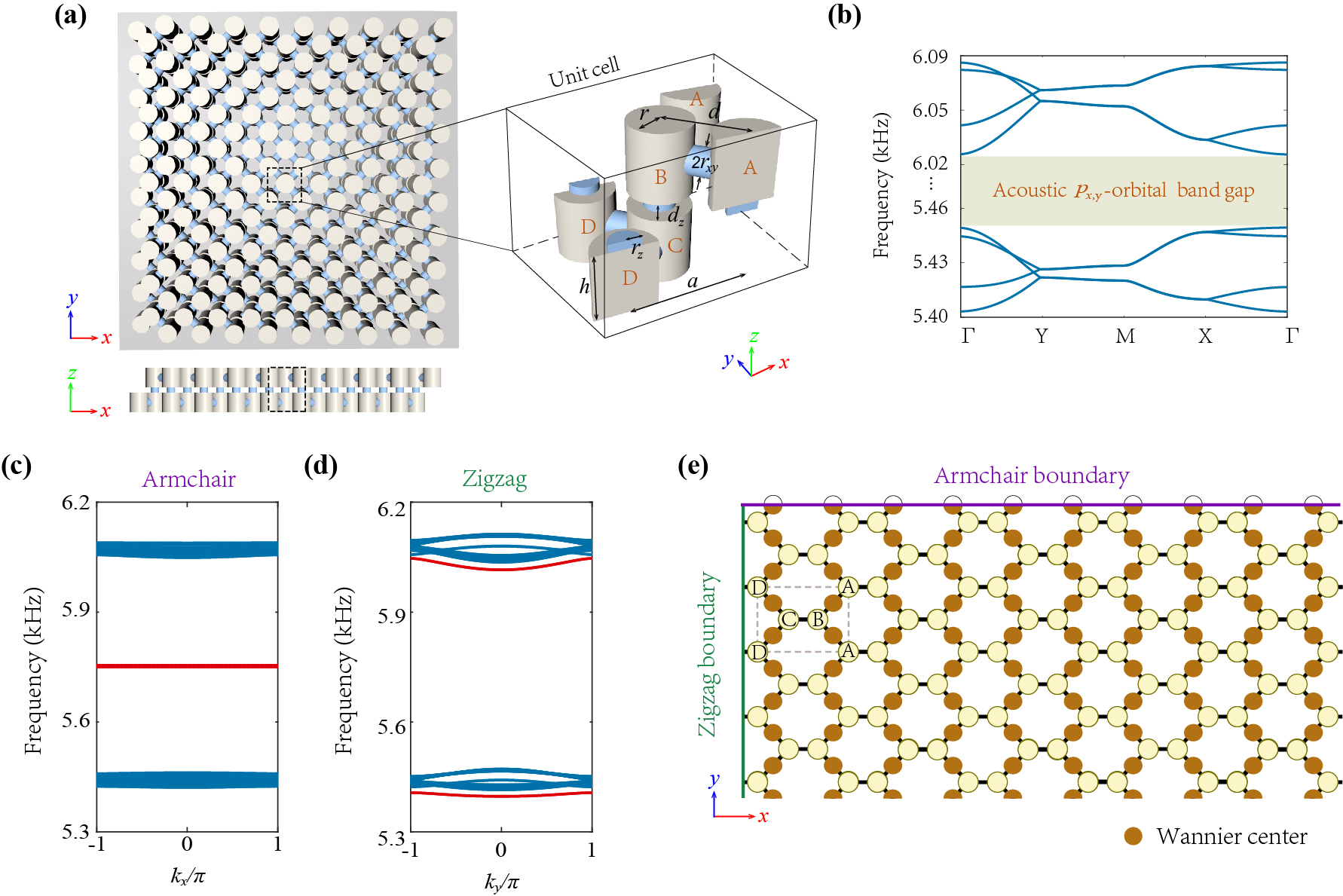}
	\caption{(a) Acoustic realization of the tight-binding phosphorene lattice. The left panels show the top and side views of the acoustic lattice. The cylindrical cavities and connecting tubes denote the atomic sites and hopping couplings, respectively. The enlarged panel shows the unit cell of the acoustic lattice. (b) The acoustic bulk band structure associated with the $p_{x,y}$-orbitals, showing a full band gap. The major geometric parameters in simulation are $a = 64.8mm$, $d_z  = 9.8 mm$, $h = 40mm$, $r = 18mm$, $d = 48.4mm$, $r_{xy} = 8.2 mm$, and $r_z = 8 mm$. (c) and (d) The acoustic projected band structures of $p_{x,y}$-orbitals for the armchair and zigzag edge terminations, respectively. The acoustic edge states are highlighted in red color. (e) The Wannier center configurations induced by the $p_{x,y}$-bands below the topological band gap. The Wannier centers at the armchair boundary are cut through by the edge boundary termination.}
	\label{Fig_3}                                                       
\end{figure*}

\begin{figure*}
	\centering
	\includegraphics[width=1.85\columnwidth]{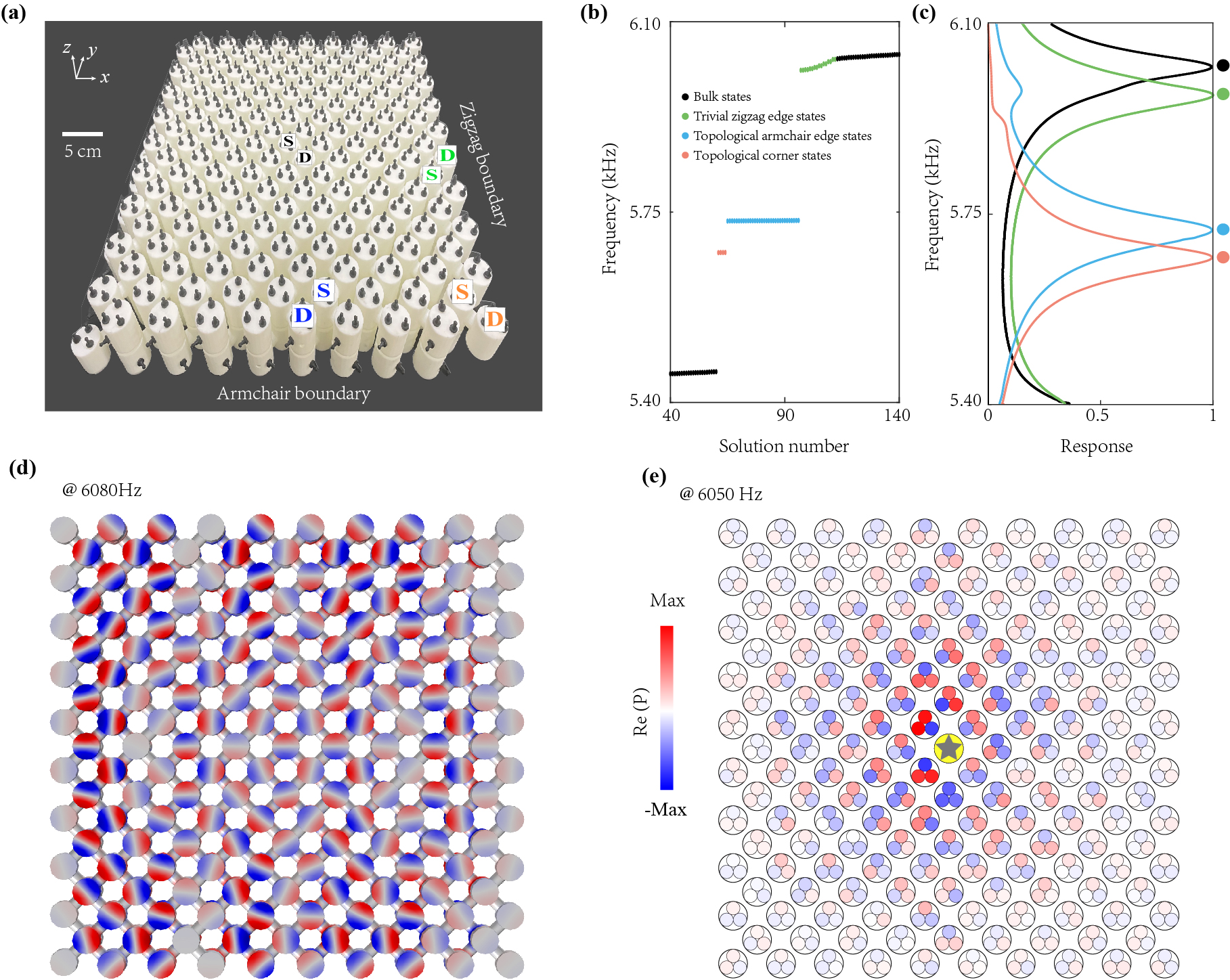}
	\caption{(a) Photograph of the experimental sample for the acoustic phosphorene lattice with $5\times5$ unit cells. The four positions labeled by $s$ with different colors denote, separately, the locations of the acoustic source to excite the bulk, the armchair edge, the zigzag edge, and the corner states. (b) Acoustic eigen-spectrum from finite-element simulation for a finite-sized acoustic metamaterial with $9\times9$ unit cells. The topological corner states (red color), armchair edge states (blue color), as well as trivial zigzag states (green color), emerge in the acoustic band gap. The black dots denote the bulk acoustic states. (c) Measured acoustic response versus frequencies corresponding to various {acoustic pump-probe} detection schemes as labeled in (a) and using the same color scheme as in (b). Each curve is normalized so that its maximum value is 1. (d) The simulated acoustic pressure profile (real part of the acoustic pressure) for the bulk states with eigenfrequency $f = 6093$ Hz. (e) Measured acoustic pressure profile (real part of the acoustic pressure) for the bulk states with excitation frequency $f = 6080$ Hz. The yellow star represents the location of the acoustic source.}
	\label{Fig_4}                                                       
\end{figure*}

\begin{figure*}[t]
    \centering
	\includegraphics[width=1.85\columnwidth]{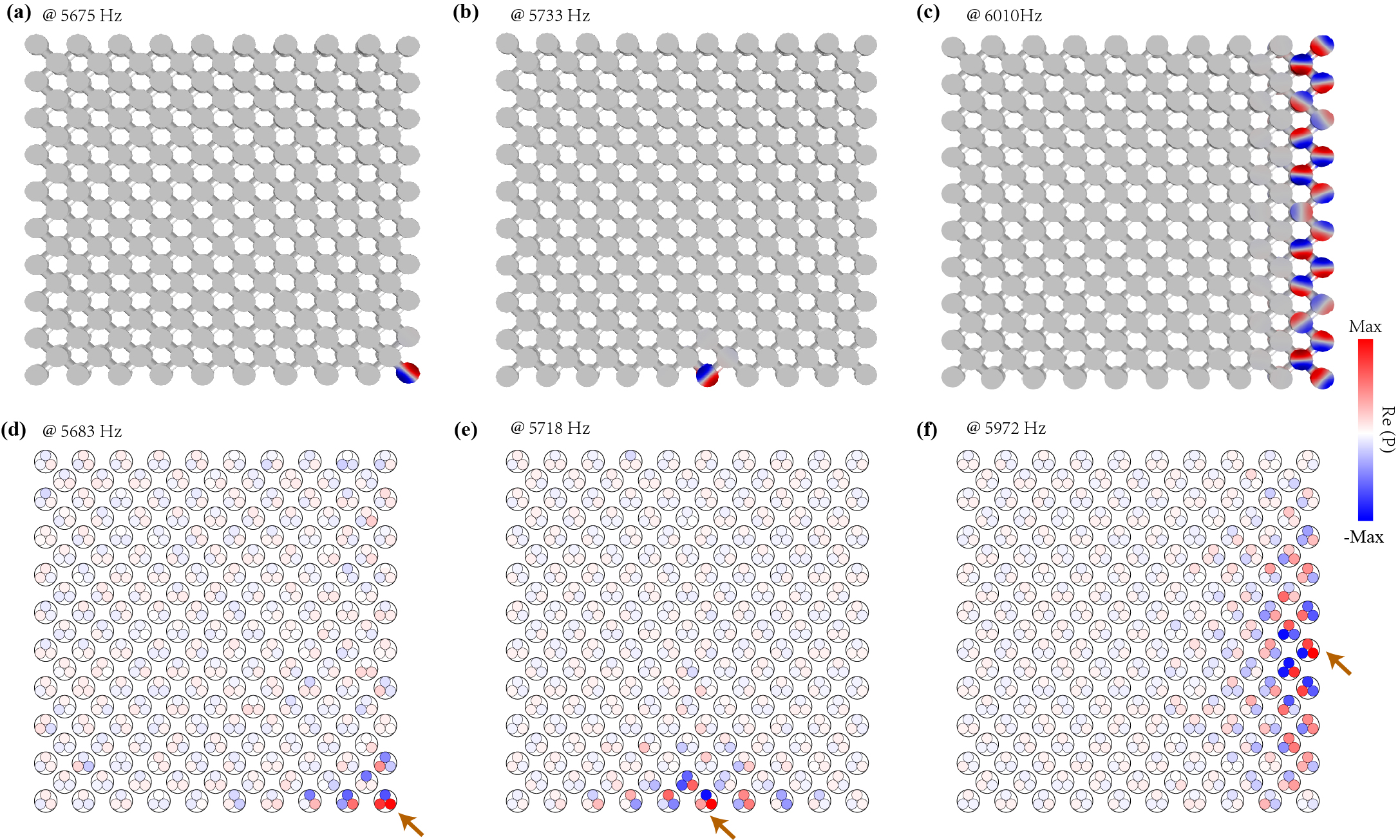}
	\caption{(a)-(c) The acoustic pressure profile (real part of the acoustic pressure) for the upper cavities from eigenstate simulations for (a) the corner state, (b) the armchair edge state, and (c) the zigzag edge state. (d)-(f) The measured acoustic pressure profiles (real part of the acoustic pressure) using the acoustic pump-probe technique for (d) the corner state, (e) the armchair edge state, and (f) the zigzag edge state. The arrows in (d)-(f) label the location of the acoustic source. The corresponding eigenfrequencies and excitation frequencies are labeled above each figure.}
	\label{Fig_5}                                                       
\end{figure*}

\subsection{Phosphorene-lattice acoustic metamaterial}\label{bb}
We use a phosphorene-lattice air-borne acoustic metamaterial to realize the above TB model. The basic approach is to use the cylindrical acoustic cavities to simulate the atomic sites that support the $s$ and $p_{x,y,z}$ orbitals, while using the connecting tubes to simulate the couplings between these orbitals. Here, as the solid walls can be treated as the acoustic hard boundaries, the couplings between the cylindrical acoustic cavities can be well engineered to realize the targeted couplings in the TB model. The right panel in Fig.~\ref{Fig_3}(a) gives a zoom-in view of a unit cell of the designed acoustic metamaterial. The acoustic waves are propagating in the (grey) acoustic cavities and the (blue) connecting tubes. We remark that the utilization of cylindrical cavities instead of spherical ones enables effective discrimination among onsite frequencies of $s$-, $p_z$, and $p_{x,y}$-orbitals, thereby facilitating the separation of the bands developed from these orbitals in frequency. Here, we focus mainly on the acoustic bands derived from the $p_{x,y}$-orbitals. The calculated bulk band structure induced by $p_{x,y}$-orbitals is depicted in Fig.~\ref{Fig_3}(b), confirming the consistency between the phosphorene lattice acoustic metamaterial and the TB model. In this study, we employ the commercial finite-element solver COMSOL Multiphysics to perform the acoustic wave simulations. In the simulations, the air density and the acoustic velocity are set as $1.29$~kg/m$^3$ and 347~m/s, respectively. The geometric parameters used in the simulations are $a = 64.8$~mm, $d_z  = 9.8$~mm, $h = 40$~mm, $r = 18$~mm, $d = 48.4$~mm, $r_{xy} = 8.2$~mm, and $r_z = 8$~mm. With such a simple design, the acoustic metamaterial can already reproduce the main features of the TB model.

It is worth mentioning that the main deviation between the acoustic metamaterial and the TB model is that in the former there are inevitable long-range acoustic couplings that break the chiral symmetry for the $p$-orbital bands. This effect leads to the shift of the frequency of the edge and corner states from the center of the band gap. Often this effect also leads to trivial edge states. In this work, thanks to the limited strength of the long-range couplings for the $p_{x,y}$ orbitals, the chiral symmetry is only slightly broken. Beside these side effects, the acoustic metamaterial realizes the main properties of the the bulk, edge, and corner states in the TB model.

The acoustic band structures obtained for the ribbon-shaped supercells with the armchair and zigzag edge boundaries are presented in Figs.~\ref{Fig_3}(c) and~\ref{Fig_3}(d), respectively. It is evident that the flat dispersion of the armchair edge states (highlighted in red) agrees well with the TB model. For the zigzag edge boundaries, due to the breakdown of the chiral symmetry in the acoustic metamaterial, edge states induced by trivial boundary effects appear in the band gap. These edge states are very close to the bulk bands. By adjusting geometric parameters, these states can be easily shifted into the bulk bands without closing the band gap due to their lack of topological protection. As the chiral symmetry is only slightly broken in our acoustic metamaterial, the armchair edge states remain to exhibit flat band which is an intriguing feature of our system. To provide a better understanding of the two types of acoustic edge states regarding their topological and trivial nature, we illustrate a top view schematic illustration of the Wannier center configurations for the $p_{x,y}$-bands below the band gap in a finite lattice in Fig.~\ref{Fig_3}(e). It is seen that the armchair edge boundary cut through the Wannier centers whereas the zigzag edge boundary does not.

\subsection{Observation of the $p_{x,y}$-orbital edge and corner states}\label{cc}
We investigate the higher-order topological phenomena of the $p_{x,y}$-orbitals in the phosphorene lattice through both simulation and experimental approaches. The experimental sample, depicted in Fig.~\ref{Fig_4}(a), consists of $5\times5$ unit cells fabricated using three-dimensional printing technology with photosensitive resin. The walls that form the cylindrical cavities and the connecting tubes have the thickness of about 2~mm. The resin acts as a hard-wall boundary due to its mismatched acoustic impedance with air, enclosing the air regions where acoustic waves can be stimulated and propagate. The geometric parameters used in experiments are identical to those used in the simulations in Fig.~\ref{Fig_3}. To facilitate measurements, each cylindrical cavity is equipped with small holes for acoustic signal excitation and detection. Specifically, considering the $p$-orbital feature of acoustic waves, we include three holes on either the top or bottom surface of each cylindrical cavity. In Fig.~\ref{Fig_4}(a), four different colored letters ``S'' represent the locations of the acoustic source that excites the bulk states, the armchair edge states, the zigzag edge states, and the corner states respectively, totaling four different pump-probe setups.

We first calculate the eigen-spectrum of the experimental sample using acoustic simulations [see Fig.~\ref{Fig_4}(b)], which reveal four degenerate corner states highlighted in red color, 32 topological armchair edge states (blue color), along with trivial zigzag edge states within the band gap and the bulk states, if we consider only the eigenstates developed from the $p_{x,y}$ orbitals. The calculated eigen-spectrum indicates the higher-order topology associated with $p_{x,y}$-bands which is the focus of this study.

It is worth noting that the number of the armchair edge states matches precisely with the number of the Wannier centers cut through by the armchair edge boundary, which confirms the topological origin from the Wannier center picture. In addition, due to the slight breaking of the chiral symmetry, the degeneracy between the corner states and the armchair edge states is lifted. It is seen from Fig.~\ref{Fig_4}(b) that the corner states now have different frequency with the armchair edge states. This effect turns out to be an advantage for the experimental probe of the corner states and the armchair edge states, since they do not mix with each other in the frequency.

To detect the spectral response around the bulk, armchair, and zigzag edges, as well as corner regions, we employ four pump-probe configurations. A tiny speaker is positioned at the location labeled by the letter ``S'' in Fig.~\ref{Fig_4}(a) and connected to an Agilent network analyzer for frequency excitation. Simultaneously, a miniature microphone captures the resulting acoustic signal in other cavities through the small holes at the top of bottom of the cylindrical cavities [denoted by the letter ``D'' in Fig.~\ref{Fig_4}(a)]. Both the speaker and the microphone are inserted into the cavities through the small holes (We keep the other unused holes closed during the measurements). As illustrated in Fig.~\ref{Fig_4}(c), the resonant frequencies corresponding to the peaks in the four pump-probe response curves exhibit excellent consistency with the simulated eigen-spectrum shown in Fig.~\ref{Fig_4}(b).

To visualize the higher-order topological phenomena, we present in the following both the simulated and the experimentally observed features of the bulk, edge, and corner states. Fig.~\ref{Fig_4}(d) gives the acoustic pressure profile corresponding to an bulk eigenstate with eigenfrequency $f = 6093$ Hz. In comparison, Fig.~\ref{Fig_4}(e) presents the measured acoustic field profile with an excitation frequency of $f = 6080$ Hz which corresponds to the peak frequency in the bulk pump-probe response curve shown in Fig.~\ref{Fig_4}(c). Both acoustic wavefunctions exhibit the features of bulk extended states. Here, the acoustic wave amplitude detected in the pump-probe measurements is less extended which is mainly because of the non-negligible dissipation and damping that limits the propagation of the excited acoustic waves. Importantly, the acoustic wave amplitude distributions in both simulation and experiments show clear features of $p_{x,y}$-orbitals in each cavity. In Fig.~\ref{Fig_5} we present the simulated eigenstates wavefunctions and the experimentally detected acoustic wave amplitudes for the corner states, the armchair edge states, and the zigzag edge states. The detected acoustic wave amplitudes are obtained from the {acoustic pump-probe measurements at the resonance frequencies for various response curves} as labeled in Fig.~\ref{Fig_4}(c).

A sharp feature of the corner states here is its $p_{x,y}$ orbital nature which is confirmed in both the simulation [Fig.~\ref{Fig_5}(a)] and the experimental results [Fig.~\ref{Fig_5}(d)]. Intriguingly, due to the flat band property of the armchair edge states, the eigenstates wavefunction of such edge states are highly localized [Fig.~\ref{Fig_5}(b)]. However, in experiments due to the fact that the {acoustic pump-probe} measurements inevitably involve evanescent signals from the source and the intrinsic acoustic dissipation, the detected acoustic wave amplitude profile cannot be localized within an cavity [This also happens for the corner state measurements]. Nevertheless, the $p_{x,y}$ orbital nature is clearly seen in the detected acoustic wavefunction for the armchair edge states [Fig.~\ref{Fig_5}(e)]. Finally, the trivial zigzag edge states can also be seen in the simulation [Fig.~\ref{Fig_5}(c)] and the experiments [Fig.~\ref{Fig_5}(f)] which also show the $p_{x,y}$ orbital features. Furthermore, compared to the trivial zigzag edge states, the topological armchair edge states exhibit stronger localization {due to their nearly flat dispersion, as indicated in Fig.~\ref{Fig_3}(c)}. Overall, the consistency between the eigenstate simulations and the {acoustic pump-probe} measurements here confirm the emergence of the topological corner and armchair edge states in real acoustic systems.

\section{Conclusion and outlook}\label{dd}
In summary, we have successfully demonstrated the existence of $p_{x,y}$-orbital-induced topological edge and corner states in an acoustic phosphorene metamaterial with puckered geometry. Remarkably, the higher-order topology found here is unique to the $p_{x,y}$-orbitals, while absent for the $s$- or $p_z$-orbitals. We use acoustic pump-probe measurements to verify the theoretical predictions. Our findings highlight the alternative mechanisms for topological phases developed from $p$-orbitals and shed light on the manipulation of acoustic waves by utilizing the orbital degrees of freedom. Furthermore, the findings in this work demonstrate the versatile role of higher orbitals in engineering phononic states and acoustic dynamics. The combination of real-space structure (e.g., the puckered lattice geometry in this work) and higher orbitals could possibly provide unconventional metamaterial solutions for application challenges. For instance, we expect that our approach can be generalized to elastic metamaterials~\cite{huang2021recent} where puckered lattice geometry and $p$-orbital effects can lead to rich physics and novel manipulation of elastic (phononic) waves. In such elastic metamaterials, the flat-band edge states can be used to enhance the nonlinear effects in the edge channel which may lead to unconventional topological phononic phenomena.

\section*{ACKNOWLEDGMENTS}
This work was supported by the National Key R\&D Program of China (2022YFA1404400), the National Natural Science Foundation of China (Grant Nos. 12125504 and 12074281), the “Hundred Talents Program” of the Chinese Academy of Sciences, and the Priority Academic Program Development (PAPD) of Jiangsu Higher Education Institutions.

\end{document}